
\magnification\magstep1

\font\rfont=cmr10 at 10 true pt
\def\ref#1{$^{\hbox{\rfont {[#1]}}}$}


\font\twelvebf=cmbx12

\def\a {\alpha}   
\def\e{\epsilon}  \def\o{\omega}
\def\s {\sigma}

\def\pmb#1{\setbox0=\hbox{#1}
 \kern.05em\copy0\kern-\wd0 \kern-.025em\raise.0433em\box0 }

\def \half {{\scriptstyle {1 \over 2}}}

 %

\def \i {\item}

\parskip=6pt
\parindent=0pt
\hsize=17truecm\hoffset=-5truemm
\voffset=-1truecm\vsize=26.5truecm
\def\footnoterule{\kern-3pt
\hrule width 17truecm \kern 2.6pt}


\catcode`\@=11 

\def\nolabels{\def\wrlabeL##1{}\def\eqlabeL##1{}\def\reflabeL##1{}}
\def\writelabels{\def\wrlabeL##1{\leavevmode\vadjust{\rlap{\smash%
{\line{{\escapechar=` \hfill\rlap{\sevenrm\hskip.03in\string##1}}}}}}}%
\def\eqlabeL##1{{\escapechar-1\rlap{\sevenrm\hskip.05in\string##1}}}%
\def\reflabeL##1{\noexpand\llap{\noexpand\sevenrm\string\string\string##1}}}
\nolabels
\global\newcount\refno \global\refno=1
\newwrite\rfile
\def\defref{$^{{\rfont [\the\refno]}}$\nref}
\def\nref#1{\xdef#1{\the\refno}\writedef{#1\leftbracket#1}%
\ifnum\refno=1\immediate\openout\rfile=refs.tmp\fi
\global\advance\refno by1\chardef\wfile=\rfile\immediate
\write\rfile{\noexpand\item{#1\ }\reflabeL{#1\hskip.31in}\pctsign}\findarg}
\def\findarg#1#{\begingroup\obeylines\newlinechar=`\^^M\pass@rg}
{\obeylines\gdef\pass@rg#1{\writ@line\relax #1^^M\hbox{}^^M}%
\gdef\writ@line#1^^M{\expandafter\toks0\expandafter{\striprel@x #1}%
\edef\next{\the\toks0}\ifx\next\em@rk\let\next=\endgroup\else\ifx\next\empty%
\else\immediate\write\wfile{\the\toks0}\fi\let\next=\writ@line\fi\next\relax}}
\def\striprel@x#1{} \def\em@rk{\hbox{}}
\def\lref{\begingroup\obeylines\lr@f}
\def\lr@f#1#2{\gdef#1{\defref#1{#2}}\endgroup\unskip}
\newwrite\lfile
{\escapechar-1\xdef\pctsign{\string\%}\xdef\leftbracket{\string\{}
\xdef\rightbracket{\string\}}}

\def\writestop{\def\writestoppt{\immediate\write\lfile{\string\p
ageno%
\the\pageno\string\startrefs\leftbracket\the\refno\rightbracket%
\string\def\string\secsym\leftbracket\secsym\rightbracket%
\string\secno\the\secno\string\meqno\the\meqno}\immediate\closeout\lfile}}
\def\writestoppt{}\def\writedef#1{}
\catcode`\@=12 

{\nopagenumbers
\rightline{CERN-TH 6635/92}
\vskip 5truecm
\centerline{{\twelvebf TOTAL CROSS SECTIONS}}
\vskip 1 truecm
\centerline{A Donnachie}

\centerline{Department of Physics, University of Manchester}
\medskip
\centerline{P V Landshoff}

\centerline{CERN, Geneva$^*$}
\footnote{}{$^*$ On leave of absence from DAMTP, University of Cambridge}
\vskip 4truecm
{\bf Abstract}

Regge theory provides a very simple and economical description of all total
cross sections
\vfill
CERN-TH 6635/92\hfil\break
September 1992
\vskip 2truecm
\eject
.\vfill
\eject}
\pageno=1
\openup 7pt
The latest Review of Particle Properties\defref\one{
Particle Data Group, Physical Review D45 (1992) number 11
}
displays plots of a number of hadronic total cross sections and quotes fits
to them of the form
$$
\sigma ^{\hbox{{\sevenrm TOT}}}
=A+BP^n+C(\log P)^2+D\log P
\eqno(1)
$$
V Flaminio has kindly made available to us computer listings of the data on
which these fits are based, and so we have taken the opportunity to make
alternative fits to the data. Our fits are of Regge type, and therefore they
have a sounder theoretical basis\defref\collins{
P D B Collins, {\it Introduction to Regge theory}, Cambridge University Press
(1977)
}.
For the 9 cross sections we consider
we use only 16 free parameters, whereas fits of the type (1) need 45,
but in the high-energy region where Regge theory is applicable ($\sqrt s$
greater than about 5 or 10 GeV) our $\chi ^2$
are much the same. Further, as we discuss, it is interesting to try to
understand why the parameters in the Regge fits take the values that they do,
 but
it is difficult to attach much meaning to the parameters in the fits (1).

Our Regge fits are a sum of two powers:
$$
\sigma ^{\hbox{{\sevenrm TOT}}}
=Xs^{\e}+Ys^{-\eta}
\eqno(2)
$$
where we identify the first term as arising from pomeron exchange and the
second
from $\rho ,\o ,f,a$ exchange. Because the pomeron has the quantum numbers of
 the
vacuum, its couplings to a particle $a$ and its antiparticle $\bar a$
are equal, so that the values of the coefficient $X$ should be equal for
$\s (ab)$ and $\s (\bar ab)$.
We emphasise that both $\e$ and $\eta$ are {\it effective}
powers: they vary slowly with $s$. However the data indicate that this
variation is very slow, and here we treat them as constant.
{}From our previous work\defref\two{
A Donnachie and P V Landshoff, Nuclear Physics B231 (1983) 189
}
we expect that the value of $\e$ should be close to 0.08 and that $\eta$ is
about $\half$.

The data are most extensive for $pp$ and $\bar pp$.
We first made a simultaneous fit to the $pp$ and $\bar pp$ data for
$\sqrt s > 10$ GeV, requiring the same values of $X$, $\e$ and $\eta$
for both reactions;
that is, we have 5 parameters for the two reactions (just half the number
required with a fit of the type (1)). This gave
$$
\e = 0.0808
$$$$
\eta = 0.4525
$$
and we use {\it the same} values of these two parameters for all our subsequent
fits. For each other pair of reactions, $ab$ and $\bar ab$,
we have three new free parameters: a common value of $X$, and a $Y$ for each.
For the other reactions, we
go down to $\sqrt s=6$ GeV with our fitting procedure, because there are
fewer available data points. (It would be very nice to have more measurements
at higher energies for a variety of beams.) The values we need for $X$
and $Y$ in each case are shown in figure 1. We have given each parameter
to high accuracy because
a small change in any one parameter has a significant effect on the $\chi ^2$
of the fit. This does not mean that the parameters are  determined to that
accuracy, because the $\chi ^2$ minimum is not very sharp and
a change in any one parameter can be compensated by changes in all the
others in such a way as not to have much effect on the $\chi ^2$.
Notice also that the quality of the data is such that the precise values
of our parameters should not be taken too seriously. This is particularly
true of the coefficient $Y$ for $\s (K^+p)$.

Our  fits are compared with the data in figures 1a to 1e.
We have a number of comments on these fits:

\i{1.} Because the $pp$ and $\bar pp$ data contain such a large number of
 points,
the $\bar pp$ measurements at the CERN collider and at the Tevatron contribute
rather little to the $\chi ^2$ per degree of freedom. So it is a definite
success of the parametrisation that it agrees with the Tevatron measurement.
Indeed we predicted in 1985\defref\three{
A Donnachie and P V Landshoff, Nuclear Physics B267 (1986) 690
}
that the cross section at $\sqrt s=1800$ GeV would be about 73 mb, which
 contrasts
with many other predictions that were 10 or more mb higher\defref\four{
M Block and R N Cahn, Physics Letters B188 (1987) 143
}

\i{2.} Notice that the rising component $Xs^{\e}$ is present already at rather
low energies, $\sqrt s=5$ GeV or so. Its form is unaffected by the onset at
 higher
energies of new production processes, such as charm or minijets. Although at
Tevatron energy there is undoubtedly a large amount of minijet production,
other
processes are reduced so as to compensate for this and there is no noticeable
overall change in the rate of rise of the total cross section. This is not
unexpected\defref\five{
M Jacob and P V Landshoff, Mod Phys Lett A1 (1987) 766
}.

\i{3.} The term $Xs^{\e}$ corresponds to pomeron exchange. A simple Regge pole
with a trajectory whose intercept is $\a (0)$ contributes a power $s^{\a
 (0)-1}$.
Our $s^{\e}$ is an effective power, with $\e$ a little less than $\a (0)-1$
because the term includes also the effect of the exchange of more than one
pomeron. These multiple exchanges make the effective power $\e$ reduce as the
energy increases\ref{\collins},
but the data indicate that it does so only very slowly and so
the effective power is almost constant.  The combined effect of the multiple
exchanges is surely rather small, otherwise we should not find that the
same effective power $\e$ fits all the different cross sections. We have
previously\ref{\three} estimated that at ISR energies the multiple exchanges
contribute only a few percent to the total cross section, but they do have an
important effect on the shape of the elastic differential cross section plotted
against $t$.
It was pointed out by Collins and
Gault\defref\six{
P D B Collins and F D Gault, Physics Letters B273 (1978) 330
}
that an almost-constant power
causes no problem with the Froissart-Martin bound, because at present
energies that bound is about 10 {\it barns}, way above the data. Indeed,
our fit $98.39s^{0.0808}$ to the rising part of $\s (\bar pp)$
exceeds $(\pi /m^2_{\pi}\log ^2(s/s_0)$, for any reasonable value
of the unknown scale $s_0$,  only when $\sqrt s>10^{24}$ GeV.  Nevertheless
we should emphasise that $\e$ is not quite constant; if we pretend that it is,
our fit predicts that the cross section at SSC energy ($\sqrt s=40$ TeV) is
120 mb, but because in reality $\e$ is a little smaller by this
energy this may be an over-estimate by a few mb.

\i{4.} Because the $Xs^{\e}$ represents pomeron exchange, which has vacuum
quantum numbers, we have constrained our fits by requiring that
the coefficient $X$
be the same for $\s (ab)$ and $\s (\bar ab)\;$\footnote{$^*$}
{This contrasts markedly with the fits of the type (1)
reported in the Review of Particle Properties\ref{\one}, which violate
the Pomeranchuk theorem rather badly and would have the $pp$ cross section
overtaking the $\bar pp$ cross section at about $\sqrt s=100$ GeV.}.
But it should also be the same for $\s (pn)$ as for $\s (pp)$. We have made a
fit with $X_{pn}=X_{\bar pn}$ not constrained by this latter
requirement and obtained the value
22.15, almost equal to $X_{pp}$. Figure 1e shows the fit with
$X_{pn}$ and $X_{\bar pn}$ constrained to be equal to $X_{pp}$.

\i{5.} We find that the ratio of the coefficients $X$ for $\pi p$ and $pp$
scattering is 0.63, close to the value 2/3 of the additive-quark rule. This
property has played an important r\^ole in the theory and the phenomenology
of the pomeron\defref\seven{
P V Landshoff and J C Polkinghorne, Nuclear Physics B32 (1971) 541\hfil\break
D G Richards, Nuclear physics B258 (1985) 267
}.
Notice that the additive-quark rule should not be exact, but should receive
corrections from hadron-radius effects\defref\eight{
P V Landshoff and O Nachtmann, Z Phys C35 (1987) 405\hfil\break
J F Gunion and D Soper, Physical Review D15 (1977) 2617
}
and, more importantly, from the fact that the $Xs^{0.0808}$ terms represent
not only single-pomeron exchange, but also include some contribution from
multiple exchanges.

\i{6.} The value of $X$ for $\s (Kp)$ is a little smaller than for $\s (\pi
p)$;
it is not understood whether this is because the pomeron coupling
to strange quarks is only about 75\% of its coupling to nonstrange quarks, or
whether instead\ref{\eight}$^,$\defref\nine{
B Povh and J H\"ufner, Physics Letters B245 (1990) 653\hfil\break

H G Dosch, E Ferreira and A Kr\"amer, preprint CERN-TH6454/92
}
it is important that the radius of the $K$ is smaller than that
of the pion. However, it is interesting that the additive-quark rule does seem
to work when hadrons carrying strangeness are involved. Since $\Sigma ^-=sdd$
and $\Omega ^-=ssd$, in the combination
$$
2\s (\Omega ^-p)-\s (\Sigma ^-p)
\eqno(3)
$$
only $su$ and $sd$ scattering is involved. Because of Zweig's rule, the only
exchange that couples both to the $s$ quark and to $u$ or $d$ is the
pomeron. Hence the additive-quark rule predicts that (3) should have
the high-energy behaviour $X_ss^{0.0808}$ with
$$
X_s=X_{pp}-3(X_{\pi p}-X_{Kp})=16.26
\eqno(4)
$$
For $\sqrt s=14$ GeV this predicts a value 24.9 mb, while experiment
gives\ref{\one} 25 $\pm$ 0.7 mb. For $\sqrt s=16$ GeV the prediction is
25.5 mb, while the data give 24.6 $\pm$ 0.7 mb.

\i{7.} Data for the $\rho ,\o ,f, a$ trajectory are shown in figure 2. The
particles in square brackets are listed in the Review of Particle
Properties\ref{\one},
but their existence remains to be confirmed.  A trajectory intercept of $\a
(0)$
would yield a power $\eta$ equal to $1-\a (0)$.
Previously\ref{\two} we fitted the trajectory by the straight line drawn in
the figure, so giving a   power
$\eta$ of 0.56, rather than the 0.4525 that we are now using.  The implied
adjustment to the line in figure 2 is perfectly acceptable, though note that
$\eta$, like $\e$, is only an effective power: it takes account not only of
$\rho ,\o ,f,a$ exchange, but also of the exchange of this trajectory combined
with that of one or more pomerons. These extra exchanges tend to reduce the
effective value of $\eta$.

\i{8.} Our fit in figure 1d to $\s (\gamma p)$ is similar to one we made
a few years ago\defref\eleven{
A Donnachie and P V Landshoff, Nuclear Physics B244 (1984) 322
}
when we considered at the same time data for low-$Q^2$ deep inelastic electron
scattering. The prediction that at $\sqrt s=200$ GeV the cross section is about
160 $\mu$b differs markedly from certain others in the literature\defref\ten{
R S Fletcher, T K Gaisser and F Halzen, Physical Review D45 (1992) 377
}
and will soon be tested at HERA.

\i{9.} The fact that all total cross sections that have been measured rise
with energy at the same rate $s^{\e}$ makes it unnatural to
attribute the rise to some
intrinsic property of the hadrons involved. It is unhelpful to adopt a
 geometrical
approach and to talk of hadrons becoming bigger and blacker as the energy
increases. Rather, the rise is a property of something that is exchanged,
the pomeron, and this is why the rise is universal. Hence, although our
analysis superficially resembles that of Cheng, Walker and T T Wu long
ago\defref\cww{
H Cheng, J K Walker and Wu, Physics Letters 44B (1973) 97
},
we do not believe that it is correct to eikonalise in the way that they
and others do.  In this our conclusions are in  accord with
the recent important results from the UA8 collaboration\defref\twelve{
UA8 collaboration: P Schlein, LPTHE'91 Conference, Geneva, July 1991,
(World Scientific, Singapore,
ed. L Di Lella) and A Brandt et al, to be submitted to Physics Letters B.
}
at the CERN collider,
which indicate that the pomeron does have a rather real existence: it can hit
hadrons hard, break them up and knock most of their fragments sharply forward.
In this respect, the pomeron resembles a hadron, and indeed UA8 have measured
 its
structure function.

\i{10.} Notice that the need for a substantial contribution to $\s (pp)$ from
$\rho ,\o ,f,a$ is in conflict with ideas about duality that were current
20 years ago.
It is tempting to try to understand the relative magnitudes of the
coefficients $Y$ for the various cross sections  by extending the
additive-quark
rule to the $\rho, \o ,f,a$ exchanges. However, if one determines their
 couplings
by using the values we have found for $Y_{pp}, Y_{\bar pp}, Y_{pn}$ and
$Y_{\bar pn}$, then the predicted values of $Y_{\pi p}$ are too large by about
50\%. A possible remedy is to assume that
the proton wave function is not
symmetric, but that rather a large fraction
of its momentum is carried by one of its
constituent $u$ quarks. This has been suggested previously for other
reasons\defref\thirteen{
I R Zhitnitsky, A A Oglobin and V L Chernyak, Sov Journ Nuc Phys 48 (1988) 536
}. We have not been able to reach any definite conclusion about this,
because one would need information about the shapes of the momentum
distributions and more accurate values of the coefficients $Y$ than the
data allow.

\i{11.} Once one has fixed the magnitudes of the coefficients $Y$ for a
pair of total cross sections, one knows the relative magnitudes of
$C=+1$ $f$ and $a$ exchange, which contributes with equal signs to both,
and $C=-1$ $\rho$ and $\o$ exchange, which contributes with opposite
signs. Inserting Regge signature factors in the standard way\ref{\collins},
we may then deduce the ratio $\rho$ of the real and imaginary parts of
the forward elastic amplitude. We (and many others) have previously made
the comparison with data for $pp$ and $\bar pp$ scattering\ref{\three}.
The only other pair of reactions for which there exist a reasonable
quantity of data for $\rho$ is $\pi ^{\pm}p$ scattering\defref\durham{
Durham data base, telnet 129.234.8.100 -- user PDG, password HEPDATA
}.  The comparison of our analysis with these data is made in figure 3.
There are also a few data points  for $\rho$ in $K^{\pm}p$ scattering.
As is seen in figure 4,  they have large errors. We have already remarked
that  $Y_{K^+p}$ is not at all well determined by
the data in figure 1c and so our calculation shown in figure 4 also has a
large uncertainty. $\rho _{K^-p}$ is particularly sensitive to $Y_{K^+p}$
and can be increased significantly by reducing the latter but still keeping
the fit to $\s (K^+p)$ within the experimental errors.

\i{12.} There exist some data\ref{\durham} for the quasi-elastic process
$$
\gamma p\to\phi p
$$
Zweig's rule has the consequence that the only exchange that couples
both to the proton and to the $\gamma p$ transition vertex is the pomeron, so
the cross section for this process should behave as $s^{2\e}/b$, where $b$
is the near-forward exponential $t$-slope. Figure 5 shows a comparison with the
data in the approximation that $b$ is constant (as both theory and experiment
suggest). It would be good to have
more data, not only for this reaction but also for $\gamma p\to J/\psi\, p$.

\bigskip

{\bf Conclusions}

Regge theory remains one of the great truths of particle physics. We have
shown how it provides an extremely simple and economical parametrisation
of all total cross sections.

\vskip 2truecm
{\it We are grateful to Gerhard Schuler for reading an early version of our
manuscript}
\vskip 2truecm
{\bf Figure captions}

1. Data for total cross-sections with fits of type (2)

2. The $\rho ,\o ,f,a$ trajectory. The line is $\a (t)=0.44+0.93t$

3. Data for the ratio of the real and imaginary parts of the forward elastic
amplitudes in $\pi p$ scattering. The curves are deduced from the fits
in figure 1b

4. Data for the ratio of the real and imaginary parts of the forward elastic
amplitudes in $Kp$ scattering. The curves are deduced from the fits
in figure 1c

5. Data for $\gamma p\to \phi p$ with curve corresponding to pomeron exchange
\vfill\eject

\medskip\immediate\closeout\rfile\writestoppt
\baselineskip=14pt{{\bf References}}\bigskip{\frenchspacing%
\parindent=20pt\escapechar=` \input refs.tmp\bigskip}\nonfrenchspacing
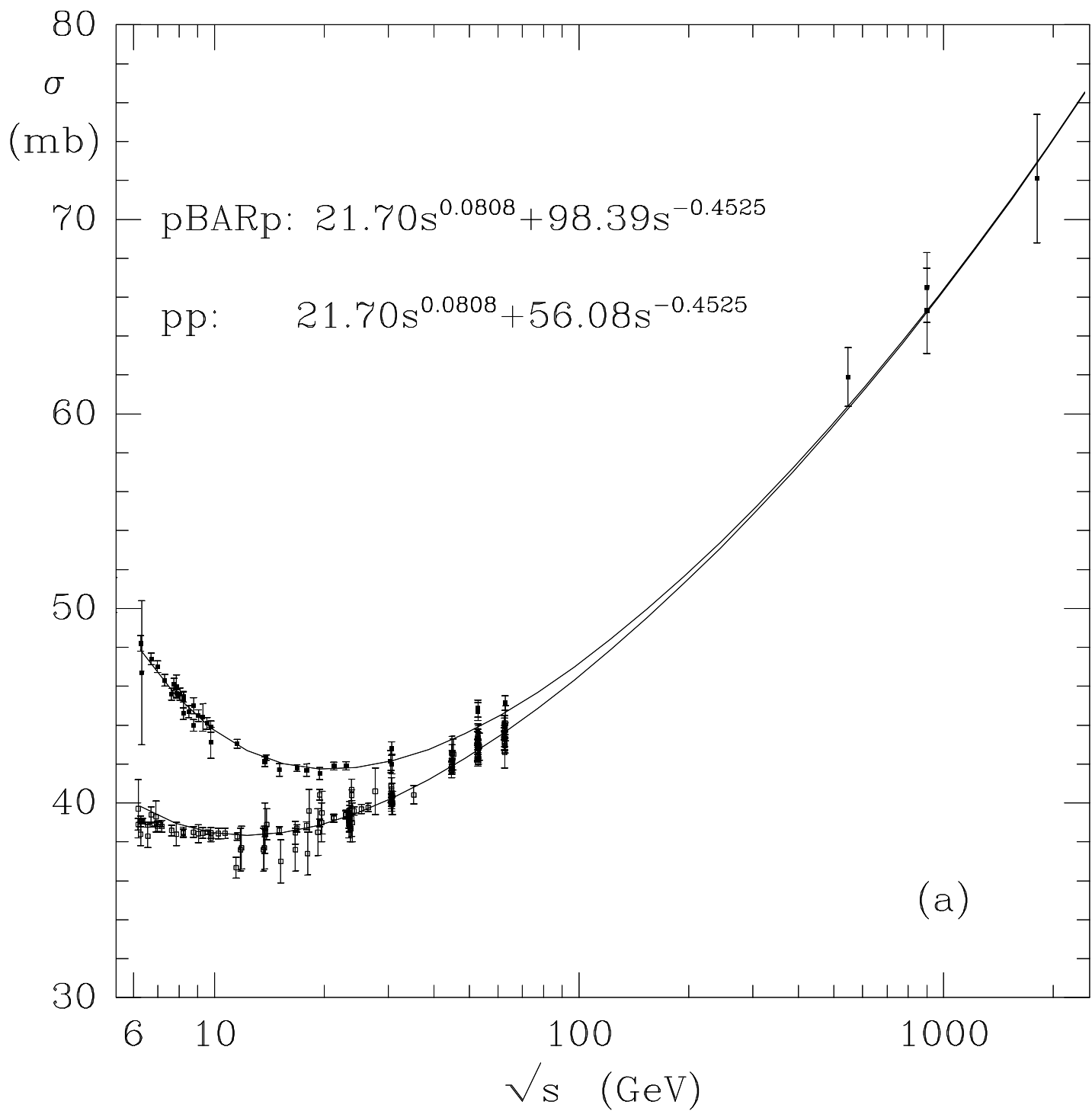
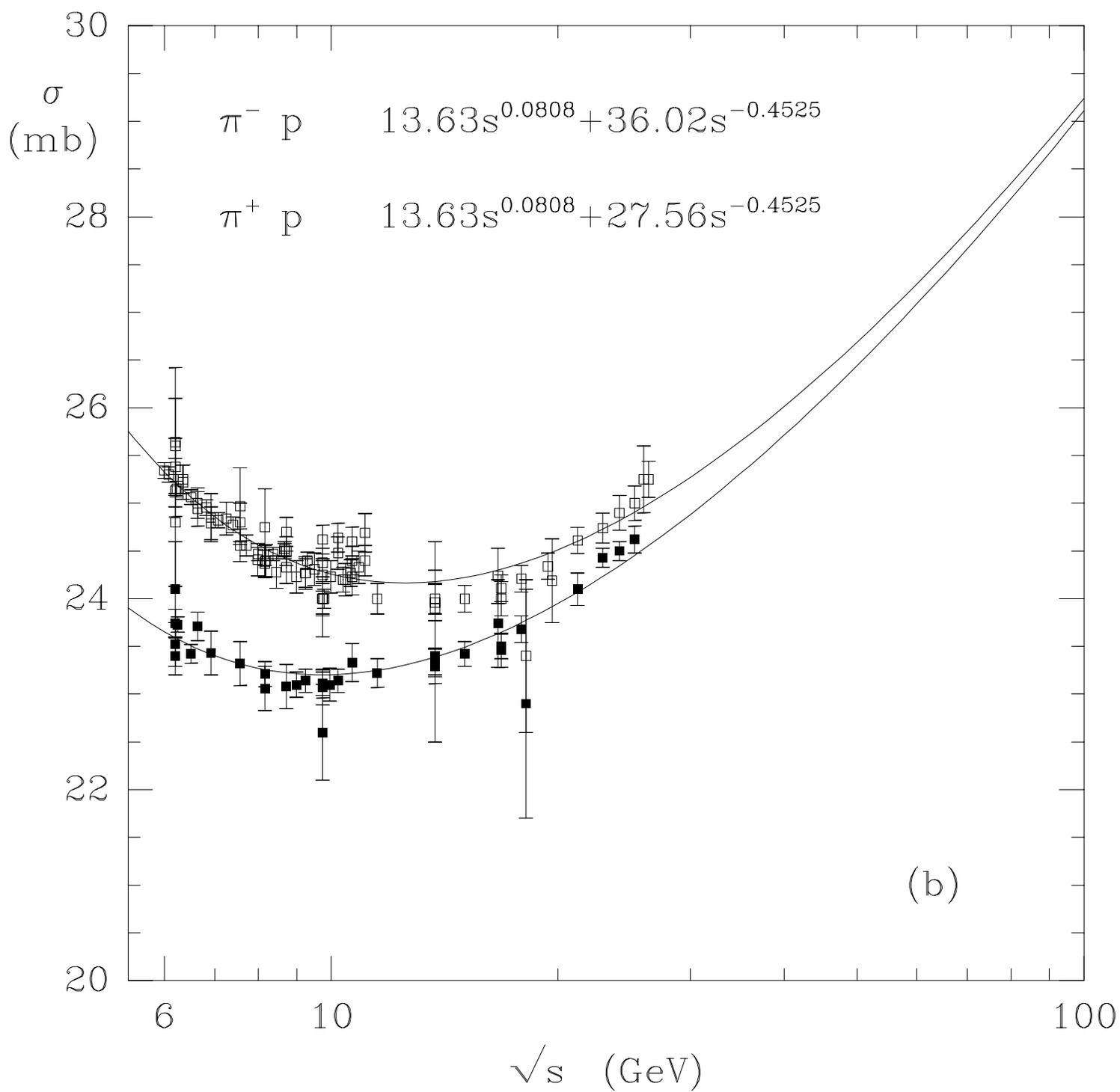
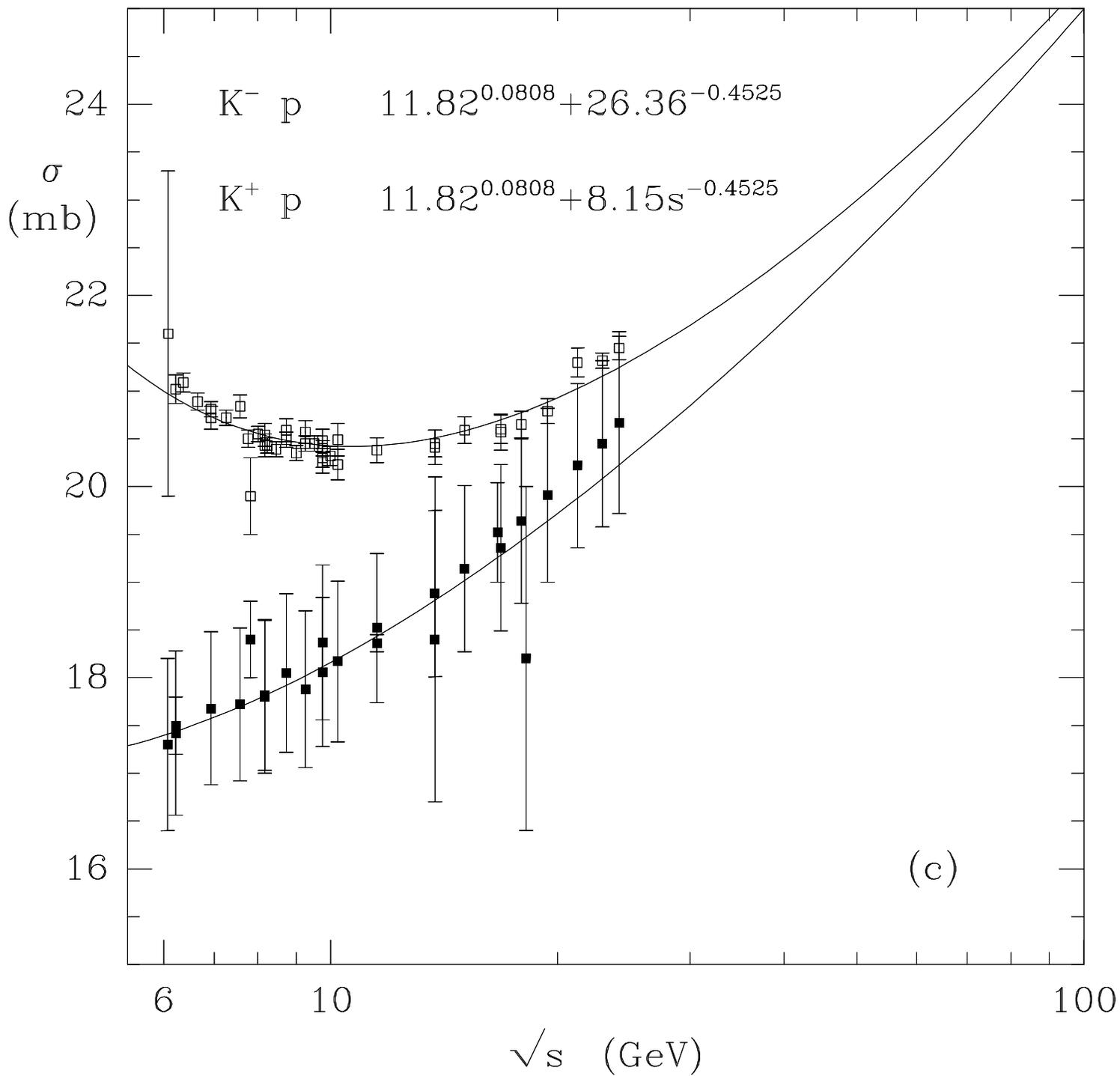
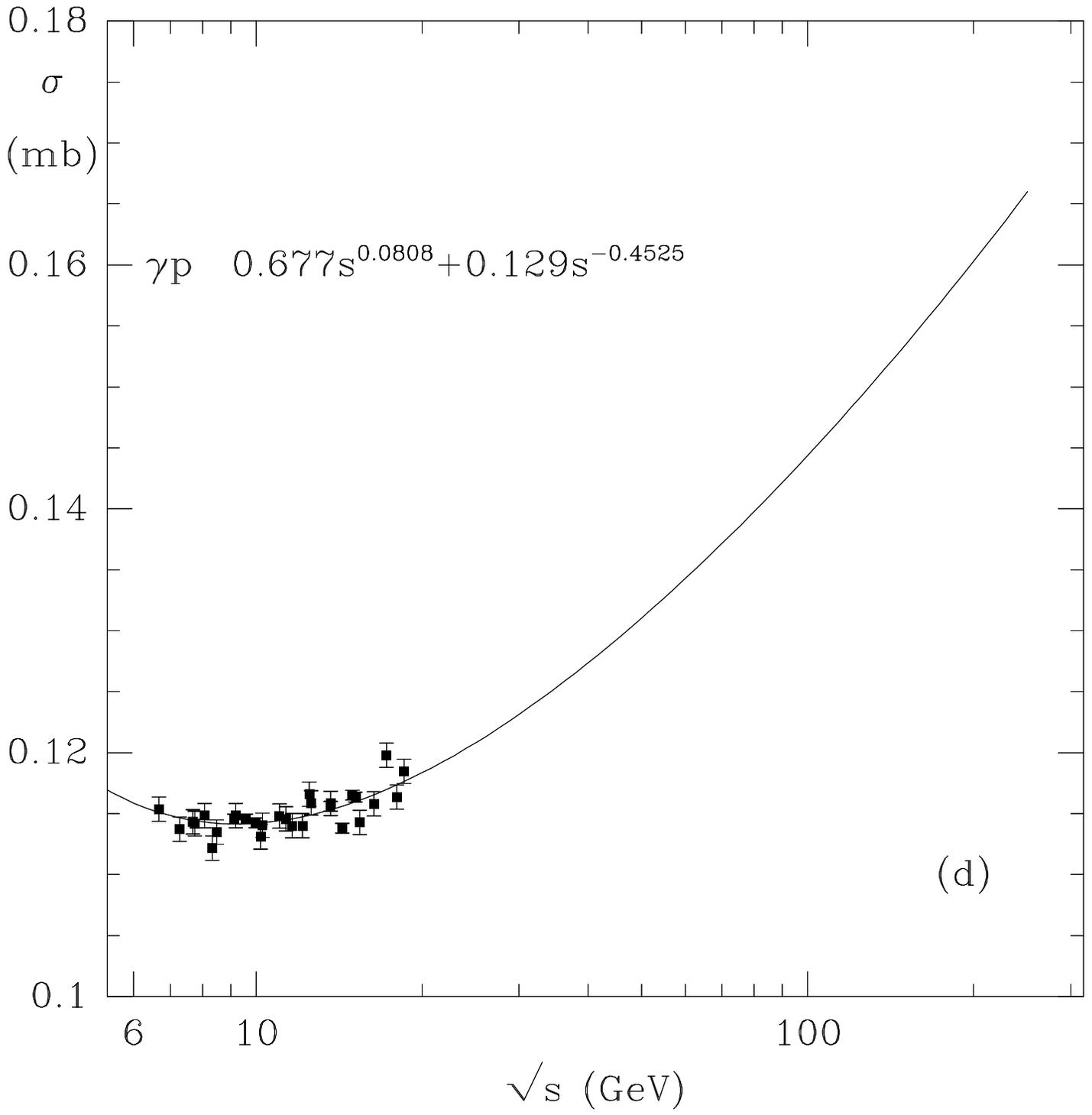
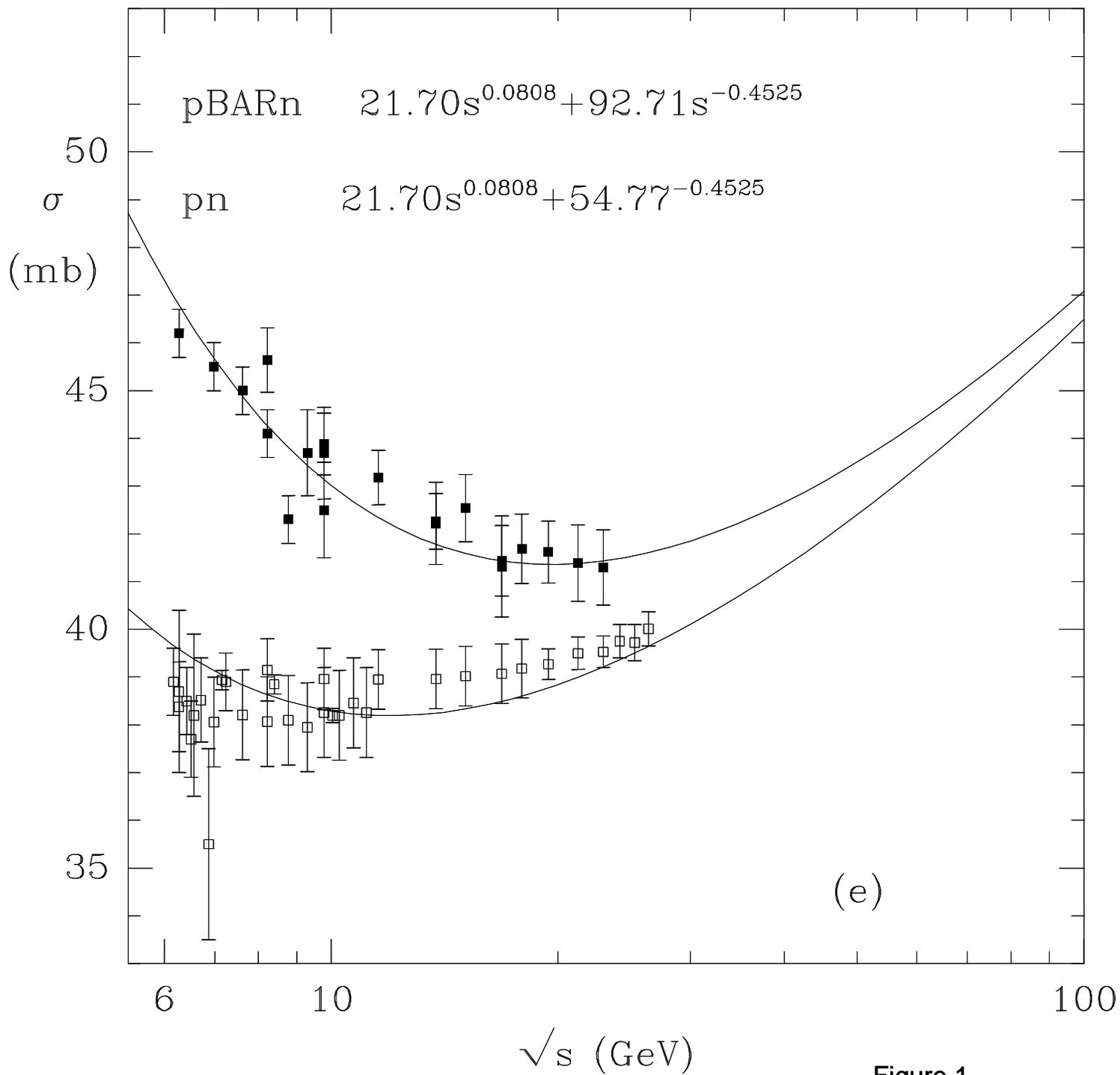
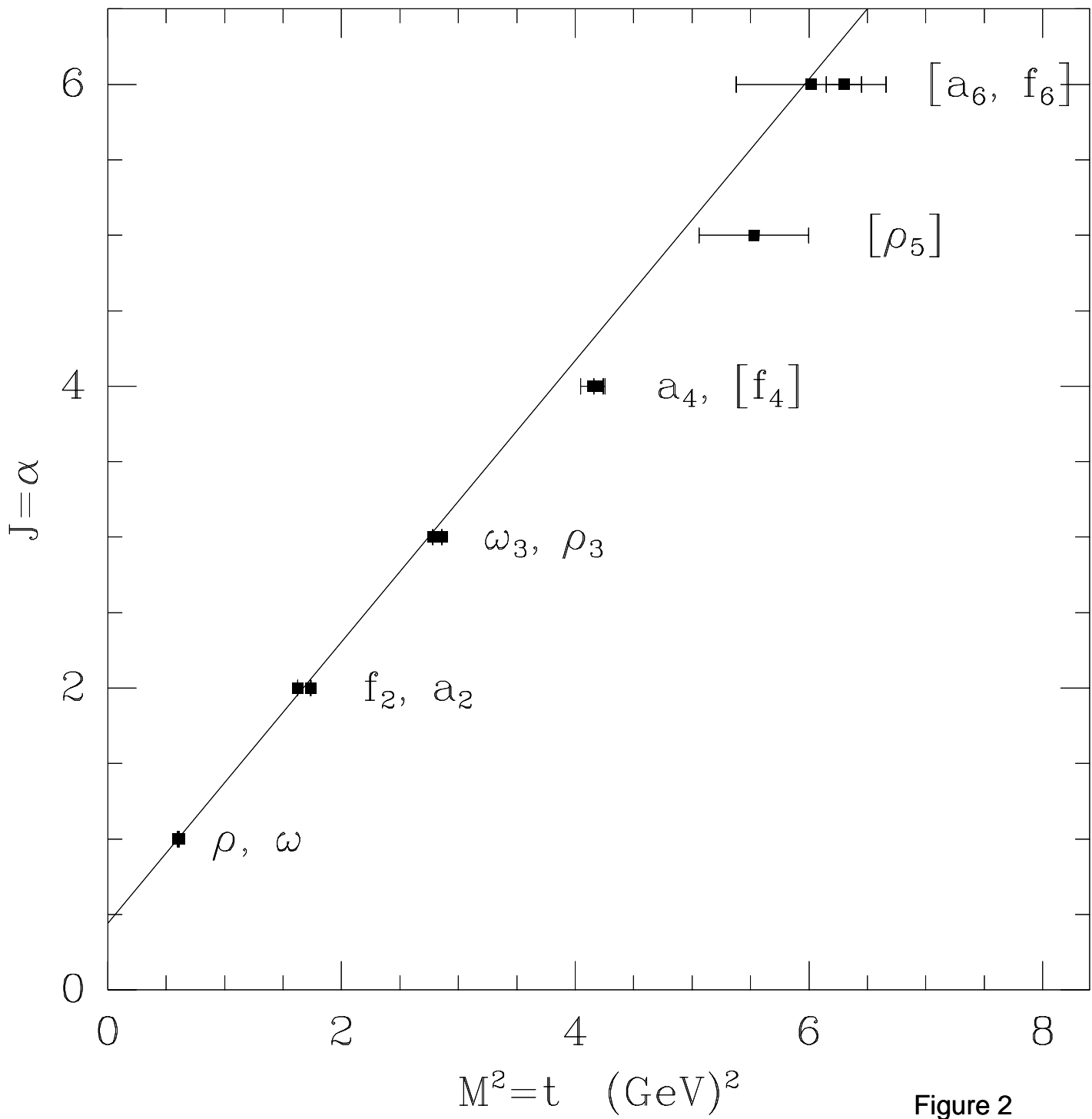
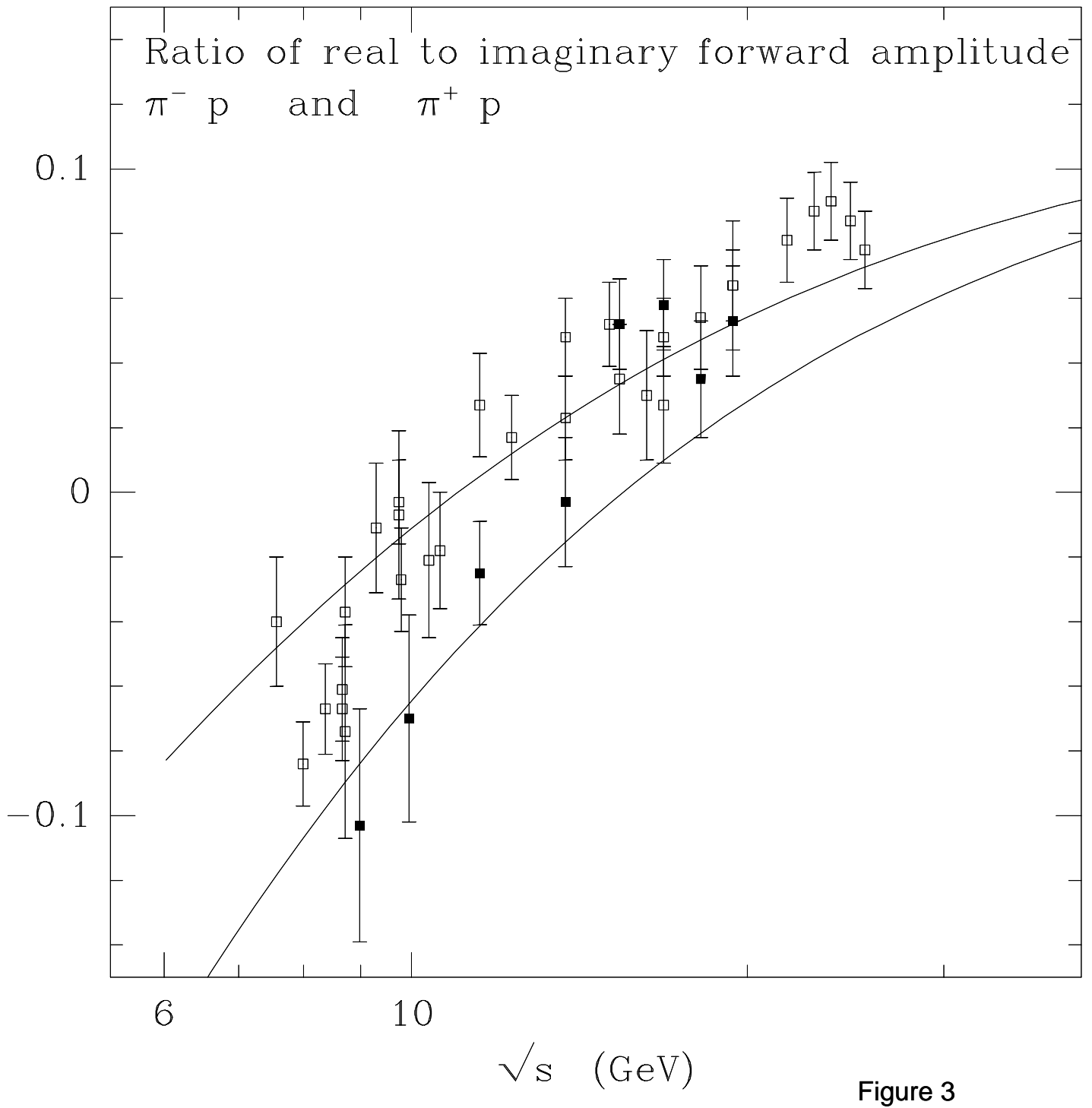
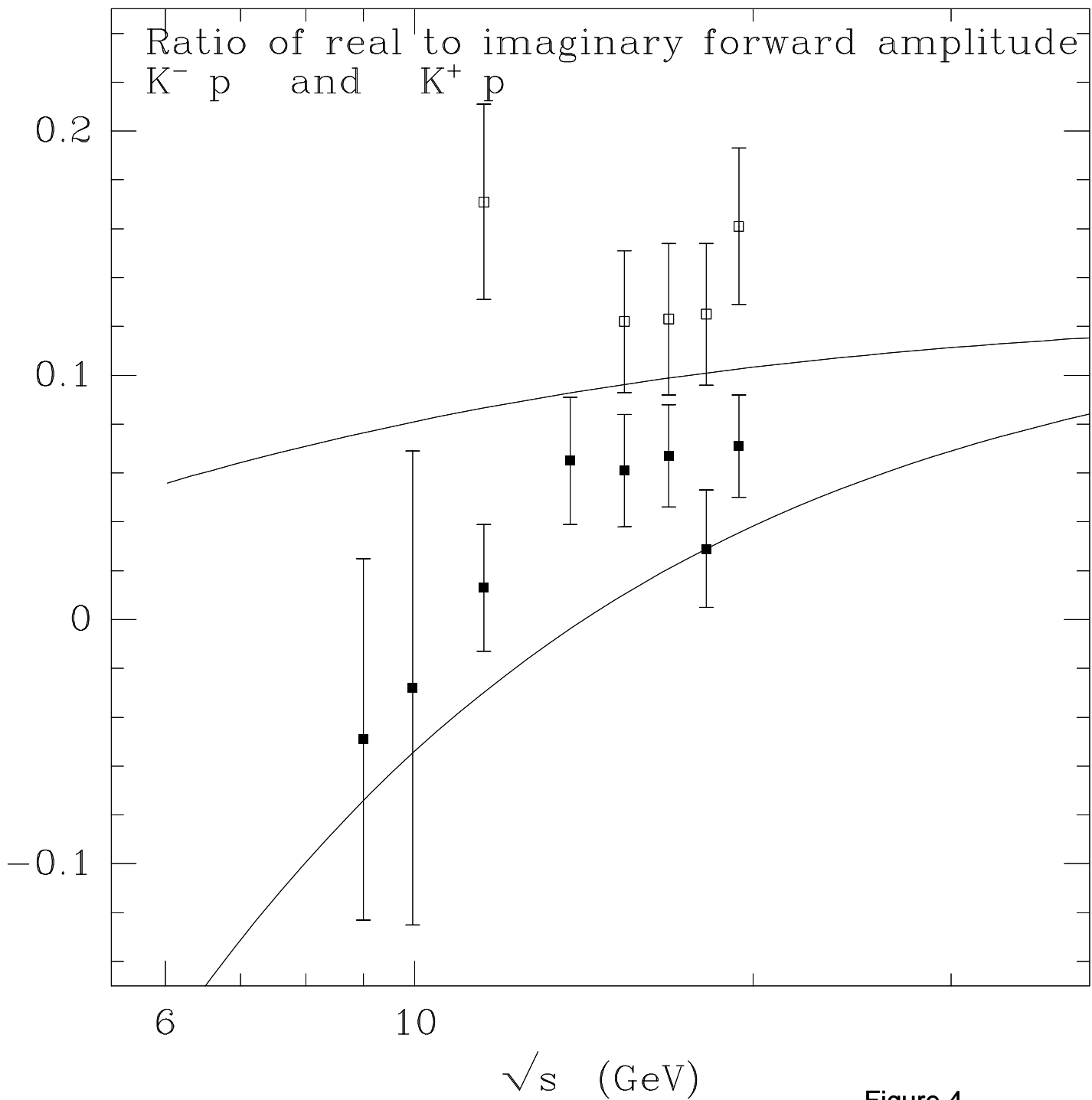
\bye